# Comment on "Particle acceleration by stimulated emission of radiation near a solid-state active medium"

Sergei Nagaitsev

Fermilab, PO Box 500, Batavia, IL 60510, USA

In a recent Letter [1], V. Berezovsky, H. Alam, and L. Schächter have reported "acceleration of electrons moving in free space near an active Nd:YAG slab". The reported mechanism is particle acceleration by stimulated emission of radiation (PASER). The energy of electrons was not measured directly, but instead inferred from measurements of the electron beam current, after it "interacted" with the Nd:YAG slab. In this Comment I suggest that the authors' proposition that the increase in beam current being due to the increase in beam energy is incorrect. Let us examine the simplified model of the experiment. The cathode emits about 5 mA of a dc electron beam. The beam is collimated by several apertures to about 100 nA before it reaches the collector, where the absorbed beam current is measured. Between the cathode and the collector this beam is "continuously illuminated" by photons from the actively pumped optical medium. After the electrons "absorb" the photons the beam current increases by 30%, to about 130 nA. Firstly, if it were true and if there were no beam collimation, this would have been the violation of charge conservation. Ignoring that fact, the authors suggest that: "Since the charge density is not expected to vary, the change in the current is due to the increase in the velocity of the electrons." This statement in itself is incorrect, as it violates the continuity of electron flux: when the beam velocity increases, its density decreases such that the beam current remains constant and the charge is conserved. Indeed, in a dc case (as described in the Letter) the continuity equation reads:

$$\frac{d(nv)}{dz} = 0, \qquad (1)$$

where $v$ is the beam velocity, $n$ is the linear charge density and $z$ is the coordinate along the beam axis. Since the dc beam current, $I = nv$ and $nv =$ const from Eq. (1), the dc beam current is not expected to vary with the change in velocity. Since the beam current in the experiment did change, it is obvious that the portion of the beam current, removed by the apertures, cannot be ignored in the analysis as it is the most likely source of the beam current increase. Moreover, there is no evidence whatsoever of the proposed PASER effect in this Letter.